\RequirePackage{amsmath}

\documentclass[runningheads]{llncs}

\usepackage{graphicx}
\usepackage{xcolor}
\usepackage{amssymb}
\usepackage{hyperref}
\usepackage{microtype}
\usepackage{booktabs}
\usepackage[linesnumbered,ruled,vlined]{algorithm2e}

\SetKwInput{KwInput}{Input}                
\SetKwInput{KwOutput}{Output} 

 2

\begin{document}

\frontmatter         

\pagestyle{headings}  

\mainmatter              

\title{On Algorithmic Estimation of Analytic Complexity for Polynomial Solutions to Hypergeometric Systems}

\titlerunning{Analytic Complexity of Solutions to Hypergeometric Systems \ldots} 

\author{V.~A.~Krasikov}

\authorrunning{V.~A.~Krasikov} 

\institute{Plekhanov Russian University of Economics \\
Laboratory of Artificial Intelligence\\
Stremyanny~36, Moscow, 115054, Russia\\
\email{Krasikov.VA@rea.ru}}

\maketitle              

\begin{abstract}
The paper deals with the analytic complexity of solutions to bivariate holonomic hypergeometric systems of the Horn type. We obtain estimates on the analytic complexity of Puiseux polynomial solutions to the hypergeometric systems defined by zonotopes. We also propose algorithms of the analytic complexity estimation for polynomials. 
\end{abstract}
\section{Introduction}

The notion of complexity is widely used in Mathematics and Computer Science in the context of several various abstract objects. The computational complexity of algorithms, the algebraic complexity of polynomials, the Rademacher complexity in the computational learning theory or the social complexity in the social systems are the concepts of great importance in the corresponding fields of science. The present work is devoted to the particular type of complexity -- the analytic complexity of bivariate holomorphic functions. 

The notion of analytic complexity is closely related to Hilbert's 13th problem, which was solved by A.N.~Kolmogorov and V.I.~Arnold in 1957~\cite{Arnold}. The initial formulation of Hilbert's 13th problem asks whether any continuous function of several variables can be represented as a finite superposition of bivariate functions~\cite{Vitushkin}. The main purpose of the theory of the analytic complexity is finding similar representations for analytic functions. The objects under consideration in this theory are the {\it analytic complexity classes.}

\begin{definition} {\rm (See~\cite{BeloshapkaRJMP2007}).
Let $\mathcal{O}(U(x_0,y_0))$ denote the set of holomorphic functions in an open neighborhood~$U(x_0,y_0)$ of a point~$(x_0,y_0)\in \mathbb{C}^2.$ The class~$Cl_0$ of analytic functions of analytic complexity zero is defined to comprise the functions that depend on at most one of the variables. A function $f(x,y)$ is said to belong to {\it the class $Cl_n$ of functions with analytic complexity $n>0$} if there exists a point $(x_0,y_0)\in \mathbb{C}^2$ and a germ~$\mathfrak{f}(x,y)\in\mathcal{O}(U(x_0,y_0))$ of this function holomorphic at $(x_0, y_0)$ such that $\mathfrak{f}(x,y)=c(a(x,y)+b(x,y))$ for some germs of holomorphic functions $a,b\in Cl_{n-1}$ and $c\in Cl_0$. If~there is no such representation for any finite~$n,$ then the function~$f$ is said to be {\it of infinite analytic complexity}}.
\end{definition}

\begin{example}
A generic element of the first complexity class~$Cl_1$ is a function of the form $f_3(f_1(x)+f_2(y)).$ A function in~$Cl_2$ can be represented in the form $f_7\left(f_5(f_1(x)+f_2(y))+f_6(f_3(x)+f_4(y))\right),$ 
\noindent where $f_i(\cdot)$ are univariate holomorphic functions, $i=1,\ldots,7$.
\end{example}

For any class of analytic complexity~$Cl_n, n\in \mathbb{N}$ there exists a system of differential polynomials with constant coefficients~$\Delta_n$ which annihilates a~function if and only if it belongs to~$Cl_n.$ 
\begin{example} (See~\cite{BeloshapkaRJMP2007}). For a bivariate function $f(x,y)$ consider the differential polynomial
$$\Delta_1 (f) = f'_{x}(f'_{y})^2f'''_{xxy}-(f'_{x})^2f'_{y}f'''_{xyy}+f''_{xy}(f'_{x})^2f''_{yy}-f''_{xy}(f'_{y})^2f''_{xx}.$$ 
This differential polynomial vanishes if and only if its argument $f\in Cl_1.$
\end{example}

The problem of defining whether a function belongs to an analytic complexity class is equivalent to computing the corresponding differential polynomial. Note that this is the problem of great computational complexity~\cite{BeloshapkaZametki2019,SadykovCASC2018}, thus a direct approach to its solution appears to be inappropriate.

An important question is a possible connection between the classes of finite analytic complexity and hypergeometric functions. In this paper we consider hypergeometric functions as solutions of hypergeometric systems in the sense of Horn~\cite{Horn1889,Sadykov-Tanabe}. We choose a matrix $A \in \mathbb{Z}^{m\times n} = \left(A_{ij},\; i=1,\ldots,m, j = 1,\ldots,n\right) $ and a vector of parameters $c = (c_1,\ldots,c_m) \in \mathbb{C}^m$. We denote the rows of this matrix by~{\bf A$_i,\, i=1,\ldots,m.$}

\begin{definition}{\rm
The {\it hypergeometric system} (or {\it Horn system}) Horn$(A,c)$ is the following system of partial differential equations:

\begin{equation}
x_j P_j(\theta) f(x) = Q_j(\theta) f(x), \, j = 1,\ldots, n,
\label{eq:HornSystem}
\end{equation}

\noindent where 

$$P_j(s)=\prod\limits_{i:A_{ij}>0}\prod\limits_{l_j^{(i)}=0}^{A_{ij}-1}\left(\left<\textbf{A}_i,s\right>+c_i+l_j^{(i)}\right), $$ 
$$Q_j(s)=\prod\limits_{i:A_{ij}<0}\prod\limits_{l_j^{(i)}=0}^{|A_{ij}|-1}\left(\left<\textbf{A}_i,s\right>+ c_i+l_j^{(i)}\right),$$ and $\theta = (\theta_1,\ldots,\theta_n),\,\theta_j=x_j\frac{\partial}{\partial x_j}.$
}
\end{definition}

\begin{definition} {\rm
The system of equations Horn$(A,c)$ is called {\it nonconfluent} if $\sum\limits_{i=1}^{m}\textbf{A}_i=0.$
}
\end{definition}

 It has been conjectured in~\cite{SadykovProceedings} that any hypergeometric function has finite analytic complexity.  Hypergeometric systems of equations differ greatly from the differential criteria for the analytic complexity classes, but numerous computer experiments suggest the hypothesis is true in a lot of particular cases~\cite{DickensteinSadykovDoklady,DickensteinSadykovMatSbornik}. The case of hypergeometric systems with low holonomic rank has been considered~in~\cite{KrasikovCASC2019}.

The set of functions of infinite analytic complexity is also a matter of interest. Until recently, all known examples of such functions were the differentially transcendental functions, that is, the functions that are not solutions to any nonzero differential polynomial with constant coefficients. Important examples of  differentially algebraic functions of infinite analytic complexity have been presented in \cite{StepanovaSbMath,StepanovaJSFU}.   

\begin{definition}{\rm
Let $l_i$ denote the generator of the sublattice $\{s\in\mathbb{Z}^n:\left<\textbf{A}_i,s\right>=0\}$ and let $k_i$ be the number of elements in the set $\{\textbf{A}_1,\ldots,\textbf{A}_m\},$ which coincide with $\textbf{A}_i.$ Let us define a polygon~$P(A)$ (see~\cite{SadykovBullSciMath}) as the integer convex polygon whose sides are translations of the vectors $k_i l_i,$ the vectors $\textbf{A}_1,\ldots,\textbf{A}_m$ being the outer normals to its sides. We will say that hypergeometric system Horn$(A,c)$ is {\it defined by the polygon}~$P(A).$
}
\label{def:OreSatoPolygon}
\end{definition}

\begin{definition} \rm
A polygon is called {\it a zonotope} if it can be represented as the Minkowski sum of segments.
\end{definition}

In this article we investigate the analytic complexity of solutions to hypergeometric systems of equations~(\ref{eq:HornSystem}) defined by zonotopes.

The present paper is organised as follows. In Section~\ref{sec:SystemsDefinedByZonotopes} we investigate particular cases of hypergeometric systems defined by zonotopes and analyze the analytic complexity of their solutions. We formulate and prove an estimate of the analytic complexity for polynomial solutions to such systems in terms of the defining matrices and parameter vectors. In Section~\ref{sec:Algorithms} we present algorithms for finding the supports of polynomial solutions to hypergeometric systems and estimating the analytic complexity of polynomials. In Section~\ref{sec:Examples} we consider examples of hypergeometric systems and estimate the analytic complexity of their solutions.

We use the Wolfram Mathematica package HyperGeometry for solving hypergeometric systems we investigate in this article. The package is available for free public use at {\small https://www.researchgate.net/publication/318986894\_HyperGeometry}, the description of available functions is given in~\cite{SadykovProgramming}. 




\section{Hypergeometric Systems Defined by Zonotopes}
\label{sec:SystemsDefinedByZonotopes}

Let us consider the special case of hypergeometric systems defined by zonotopes. Numerous experiments suggest that the analytic complexity of polynomial solutions to such systems can be much lower than its estimate based on the number of their monomials.

The set of hypergeometric systems defined by zonotopes enjoys the following properties:

\noindent a) these systems are holonomic for the generic parameter value;

\noindent b) the holonomic rank of hypergeometric systems (see Theorem~2.5 in~\cite{DickensteinSadykovAdvances}) is given by 

$$\textrm{rank}(\textrm{Horn}(A,c))=d_1 d_2-\sum\limits_{\textbf{A}_i, \textbf{A}_j \textrm{ lin. dependent}}\nu_{ij},$$ where $d_j=\sum\limits_{\scriptsize\begin{array}{c} i=1\\A_{ij}>0\end{array}}^m A_{ij}, j=1,2$ and

$$\nu_{ij}=\left\{\begin{array}{l}\textrm{min}(|A_{i1}A_{j2}|,|A_{j1}A_{i2}|), \textrm{ if } \textbf{A}_i, \textbf{A}_j \textrm{ are in opposite open quadrants of }\mathbb{Z}^2,\\0, \textrm{ otherwise.}\end{array}\right.$$

For the hypergeometric systems defined by zonotopes there is another formula for computing their holonomic rank (see Proposition~1 in~\cite{KrasikovCASC2019}), which in some cases may be more suitable;

\noindent c) the rows of the matrix defining such a system can be united into two matrices $\hat{A},-\hat{A};$ 


\noindent d) for a hypergeometric system defined by a~zonotope one can always choose parameter values such that any solution to the resulting system is a polynomial (see~\cite{Sadykov-Tanabe}). Namely, for such a hypergeometric system~$\text{Horn}(A,c),$ where the matrix $A$ contains~$2k$ rows, let $\alpha = (\alpha_1,\ldots,\alpha_k)$ be a part of the parameter vector~$c,$ corresponding to the matrix $\hat{A}$ (see the property (c) above)$, \beta = (\beta_1,\ldots,\beta_k)$~be a part of this vector, corresponding to~$-\hat{A}.$ Then the general solution to $\text{Horn}(A,c)$ is a polynomial if $-\alpha_i-\beta_i\in \mathbb{N}\backslash\{0\}$ for~$i=1,\ldots,k.$ 

The simplest case of a zonotope is a parallelogram. The analytic complexity estimate of the solutions to the systems defined by parallelograms is the basis for more complex cases.  

\begin{proposition} \rm

The analytic complexity of a hypergeometric systems defined by a parallelogram cannot exceed~$2.$

\noindent{\it Proof.}
The solution to the hypergeometric system defined by a parallelogram, has been described in Proposition~4.7 in~\cite{Sadykov-Tanabe}. For the bivariate system ($n=2$) this formula leads to $$(x_1^{-a_{11}}x_2^{-a_{21}})^{\alpha_1} \left(1+x_1^{-a_{11}}x_2^{-a_{21}}\right)^{-\alpha_1-\beta_1}\cdot (x_1^{-a_{12}}x_2^{-a_{22}})^{\alpha_2} \left(1+x_1^{-a_{12}}x_2^{-a_{22}}\right)^{-\alpha_2-\beta_2},$$where $A^{-1}=\left(\begin{array}{rr}a_{11} & a_{12} \\ a_{21} & a_{22}\end{array}\right), c=(\alpha_1,\alpha_2,\beta_1,\beta_2).$ The monomials $x_1^{-a_{11}}x_2^{-a_{21}}$ and $x_1^{-a_{12}}x_2^{-a_{22}}$ both belong to~$Cl_1,$ thus for any univariate analytic functions $\phi(\cdot),\psi(\cdot)$ the product $\phi(x_1^{-a_{11}}x_2^{-a_{21}})\cdot\psi(x_1^{-a_{12}}x_2^{-a_{22}})$ belongs to~$Cl_2.$~\hfill $\square$
\label{prop:parallelogram}
\end{proposition}

The following example shows that the solutions to hypergeometric systems defined by more complex polygons can be of a low analytic complexity.

\begin{example} {\it Simple zonotope.}
Let us consider the hypergeometric system $\text{Horn}(A_0,$ $c_0)$ defined by the matrix $A_0 = \left(\begin{array}{rrrrrr} 1 & -1 & 1 & -1 & 0 & 0 \\ 1 & -1 & 0 & 0 & 1 & -1 \end{array}\right)^{\rm T}$ and the parameter vector $c_0=(-23,22,-10,0,-9,0).$ The holonomic rank of this system is equal to~$3.$ The hypergeometric system $\text{Horn}(A_0,c_0)$ is defined by a zonotope, since rows of $A_0$ correspond to normal vectors to sides of the polygon. Representation of this zonotope in the form of the Minkowski sum of segments is shown in Figure~\ref{fig:polyOreSatoCoef}

\begin{figure}[htbp]
\begin{minipage}{4cm}
\begin{picture}(100,100)
  \put(20,0){\vector(0,1){100}}
  \put(0,20){\vector(1,0){100}}
 


  \put(20,45){\line(1,-1){25}}
  \put(45,70){\line(1,-1){25}}
  \put(20,70){\line(1,0){25}}
  \put(70,45){\line(0,-1){25}}
  \put(20,70){\circle*{3}}
  \put(45,20){\circle*{3}}  
  \put(70,45){\circle*{3}}
  \put(45,70){\circle*{3}}
  \put(20,45){\circle*{3}}
  \put(70,20){\circle*{3}}

\end{picture}
\end{minipage}
=
\begin{minipage}{2cm}
\begin{picture}(70,70)
 


  \put(10,35){\line(1,-1){25}}
  \put(35,10){\circle*{3}}  
  \put(10,35){\circle*{3}}

\end{picture}
\end{minipage}
+
\begin{minipage}{2cm}
\begin{picture}(70,70)
 


  \put(10,35){\line(1,0){25}}
  \put(35,35){\circle*{3}}  
  \put(10,35){\circle*{3}}

\end{picture}
\end{minipage}
+
\begin{minipage}{2cm}
\begin{picture}(70,70)
 


  \put(35,10){\line(0,1){25}}
  \put(35,10){\circle*{3}}  
  \put(35,35){\circle*{3}}

\end{picture}
\end{minipage}
\caption{Polygon, defining the system $\text{Horn}(A_0,c_0),$ and its representation as the Minkowski sum of segments}
\label{fig:polyOreSatoCoef}
\end{figure}
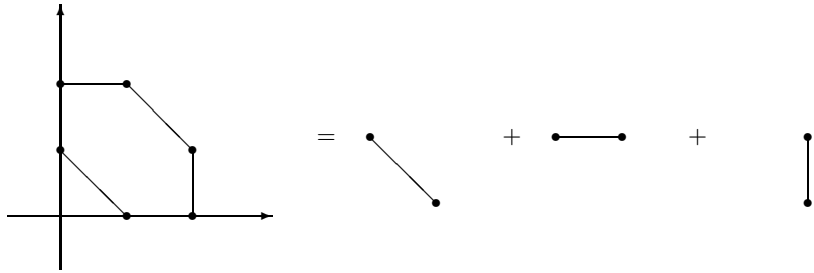

 The support for the system $\text{Horn}(A_0,c_0)$ polynomial solution is shown in Figure~\ref{fig:generalZonotopeExample}. 
\begin{figure}[htbp]
\centering
\begin{minipage}{8cm}
\begin{picture}(250,250)
  \put(20,0){\vector(0,1){260}}
  \put(0,20){\vector(1,0){260}}
  \put(8,8){\small $0$}
  \put(30,20){\line(0,1){3}}
  \put(40,20){\line(0,1){3}}
  \put(50,20){\line(0,1){3}}
  \put(60,20){\line(0,1){3}}
  \put(70,20){\line(0,1){5}}
  \put(68,8){\small $5$}
  \put(80,20){\line(0,1){3}}
  \put(90,20){\line(0,1){3}}
  \put(100,20){\line(0,1){3}}
  \put(110,20){\line(0,1){3}}
  \put(120,20){\line(0,1){5}}
  \put(120,8){\small $10$}
  \put(130,20){\line(0,1){3}}
  \put(140,20){\line(0,1){3}}
  \put(150,20){\line(0,1){3}}
  \put(160,20){\line(0,1){3}}
  \put(170,20){\line(0,1){5}}
  \put(166,8){\small $15$}
  \put(180,20){\line(0,1){3}}
  \put(190,20){\line(0,1){3}}
  \put(200,20){\line(0,1){3}}
  \put(210,20){\line(0,1){3}}
  \put(220,20){\line(0,1){5}}
  \put(216,8){\small $20$}
  \put(230,20){\line(0,1){3}}
  \put(240,20){\line(0,1){3}}
  \put(250,20){\line(0,1){3}}
  \put(20,30){\line(1,0){3}}
  \put(20,40){\line(1,0){3}}  
  \put(20,50){\line(1,0){3}}
  \put(20,60){\line(1,0){3}}  
  \put(20,70){\line(1,0){5}}
  \put(8,67){\small $5$}
  \put(20,80){\line(1,0){3}}  
  \put(20,90){\line(1,0){3}}  
  \put(20,100){\line(1,0){3}}  
  \put(20,110){\line(1,0){3}}  
  \put(20,120){\line(1,0){5}}
  \put(8,117){\small $10$}
  \put(20,130){\line(1,0){3}}   
  \put(20,140){\line(1,0){3}} 
  \put(20,150){\line(1,0){3}} 
  \put(20,160){\line(1,0){3}} 
  \put(20,170){\line(1,0){5}} 
  \put(8,167){\small $15$}
  \put(20,180){\line(1,0){3}}   
  \put(20,190){\line(1,0){3}} 
  \put(20,200){\line(1,0){3}} 
  \put(20,210){\line(1,0){3}} 
  \put(20,220){\line(1,0){5}} 
  \put(8,217){\small $20$}
  \put(20,230){\line(1,0){3}} 
  \put(20,240){\line(1,0){3}} 
  \put(20,250){\line(1,0){3}}
  \put(0,110){\line(1,0){260}}
  \put(120,0){\line(0,1){260}}
  \put(10,250){\line(1,-1){240}}
  \put(10,260){\line(1,-1){250}}
  \put(20,240){\circle*{4}}
  \put(20,250){\circle*{4}}
  \put(30,230){\circle*{4}}
  \put(30,240){\circle*{4}}
  \put(40,220){\circle*{4}}
  \put(40,230){\circle*{4}}
  \put(50,210){\circle*{4}}
  \put(50,220){\circle*{4}}
  \put(60,200){\circle*{4}}
  \put(60,210){\circle*{4}}
  \put(70,190){\circle*{4}}
  \put(70,200){\circle*{4}}
  \put(80,180){\circle*{4}}
  \put(80,190){\circle*{4}}
  \put(90,170){\circle*{4}}
  \put(90,180){\circle*{4}}
  \put(100,160){\circle*{4}}
  \put(100,170){\circle*{4}}
  \put(110,150){\circle*{4}}
  \put(110,160){\circle*{4}}
  \put(120,140){\circle*{4}}
  \put(120,150){\circle*{4}}

  \put(150,110){\circle*{4}}
  \put(160,110){\circle*{4}}
  \put(160,100){\circle*{4}}
  \put(170,100){\circle*{4}}
  \put(170,90){\circle*{4}}
  \put(180,90){\circle*{4}}
  \put(180,80){\circle*{4}}
  \put(190,80){\circle*{4}}
  \put(190,70){\circle*{4}}
  \put(200,70){\circle*{4}}
  \put(200,60){\circle*{4}}
  \put(210,60){\circle*{4}}
  \put(210,50){\circle*{4}}
  \put(220,50){\circle*{4}}
  \put(220,40){\circle*{4}}
  \put(230,40){\circle*{4}}
  \put(230,30){\circle*{4}}
  \put(240,30){\circle*{4}}
  \put(240,20){\circle*{4}}
  \put(250,20){\circle*{4}}

  \put(20,20){\circle*{4}}
  \put(30,20){\circle*{4}}
  \put(40,20){\circle*{4}}
  \put(50,20){\circle*{4}}  
  \put(60,20){\circle*{4}}
  \put(70,20){\circle*{4}}
  \put(80,20){\circle*{4}}
  \put(90,20){\circle*{4}}
  \put(100,20){\circle*{4}}
  \put(110,20){\circle*{4}}
  \put(120,20){\circle*{4}}
  \put(20,30){\circle*{4}}
  \put(30,30){\circle*{4}}
  \put(40,30){\circle*{4}}
  \put(50,30){\circle*{4}}  
  \put(60,30){\circle*{4}}
  \put(70,30){\circle*{4}}
  \put(80,30){\circle*{4}}
  \put(90,30){\circle*{4}}
  \put(100,30){\circle*{4}}
  \put(110,30){\circle*{4}}
  \put(120,30){\circle*{4}}
  \put(20,40){\circle*{4}}
  \put(30,40){\circle*{4}}
  \put(40,40){\circle*{4}}
  \put(50,40){\circle*{4}}  
  \put(60,40){\circle*{4}}
  \put(70,40){\circle*{4}}
  \put(80,40){\circle*{4}}
  \put(90,40){\circle*{4}}
  \put(100,40){\circle*{4}}
  \put(110,40){\circle*{4}}
  \put(120,40){\circle*{4}}
  \put(20,50){\circle*{4}}
  \put(30,50){\circle*{4}}
  \put(40,50){\circle*{4}}
  \put(50,50){\circle*{4}}  
  \put(60,50){\circle*{4}}
  \put(70,50){\circle*{4}}
  \put(80,50){\circle*{4}}
  \put(90,50){\circle*{4}}
  \put(100,50){\circle*{4}}
  \put(110,50){\circle*{4}}
  \put(120,50){\circle*{4}}
  \put(20,60){\circle*{4}}
  \put(30,60){\circle*{4}}
  \put(40,60){\circle*{4}}
  \put(50,60){\circle*{4}}  
  \put(60,60){\circle*{4}}
  \put(70,60){\circle*{4}}
  \put(80,60){\circle*{4}}
  \put(90,60){\circle*{4}}
  \put(100,60){\circle*{4}}
  \put(110,60){\circle*{4}}
  \put(120,60){\circle*{4}}
  \put(20,70){\circle*{4}}
  \put(30,70){\circle*{4}}
  \put(40,70){\circle*{4}}
  \put(50,70){\circle*{4}}  
  \put(60,70){\circle*{4}}
  \put(70,70){\circle*{4}}
  \put(80,70){\circle*{4}}
  \put(90,70){\circle*{4}}
  \put(100,70){\circle*{4}}
  \put(110,70){\circle*{4}}
  \put(120,70){\circle*{4}}
  \put(20,80){\circle*{4}}
  \put(30,80){\circle*{4}}
  \put(40,80){\circle*{4}}
  \put(50,80){\circle*{4}}  
  \put(60,80){\circle*{4}}
  \put(70,80){\circle*{4}}
  \put(80,80){\circle*{4}}
  \put(90,80){\circle*{4}}
  \put(100,80){\circle*{4}}
  \put(110,80){\circle*{4}}
  \put(120,80){\circle*{4}}
  \put(20,90){\circle*{4}}
  \put(30,90){\circle*{4}}
  \put(40,90){\circle*{4}}
  \put(50,90){\circle*{4}}  
  \put(60,90){\circle*{4}}
  \put(70,90){\circle*{4}}
  \put(80,90){\circle*{4}}
  \put(90,90){\circle*{4}}
  \put(100,90){\circle*{4}}
  \put(110,90){\circle*{4}}
  \put(120,90){\circle*{4}}
  \put(20,100){\circle*{4}}
  \put(30,100){\circle*{4}}
  \put(40,100){\circle*{4}}
  \put(50,100){\circle*{4}}  
  \put(60,100){\circle*{4}}
  \put(70,100){\circle*{4}}
  \put(80,100){\circle*{4}}
  \put(90,100){\circle*{4}}
  \put(100,100){\circle*{4}}
  \put(110,100){\circle*{4}}
  \put(120,100){\circle*{4}}
  \put(20,110){\circle*{4}}
  \put(30,110){\circle*{4}}
  \put(40,110){\circle*{4}}
  \put(50,110){\circle*{4}}  
  \put(60,110){\circle*{4}}
  \put(70,110){\circle*{4}}
  \put(80,110){\circle*{4}}
  \put(90,110){\circle*{4}}
  \put(100,110){\circle*{4}}
  \put(110,110){\circle*{4}}
  \put(120,110){\circle*{4}}
\end{picture}
\end{minipage}
\caption{The support for the solution of the system $\text{Horn}(A_0,c_0)$}
\label{fig:generalZonotopeExample}
\end{figure}
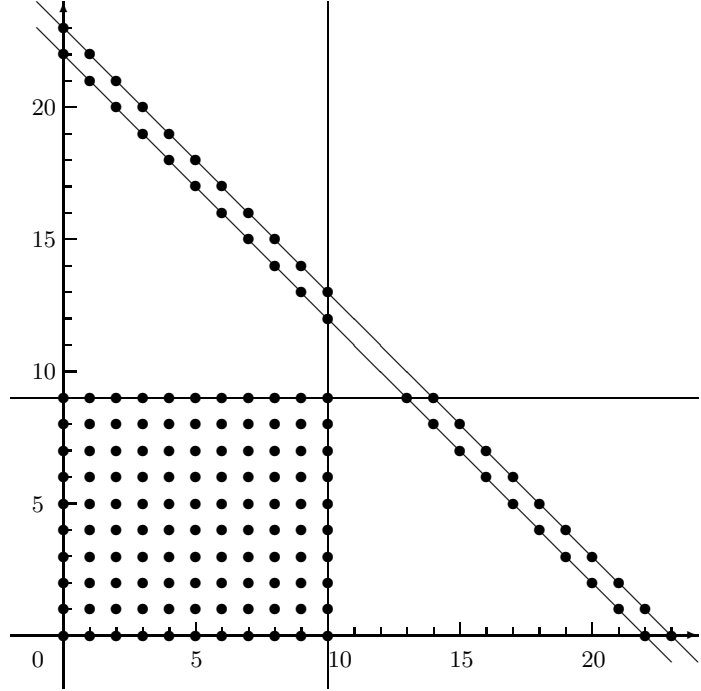
Let us consider the part of the solution whose support is bounded by the divisors parallel to coordinate axis. This polynomial~$p_0(x,y)$ belongs to the basis of the linear space of solutions to~$\text{Horn}(A_0,c_0)$. Note that $p_0(x,y)$ contains $90$ monomials (we do not put here the whole expression due to its large size) and the known estimates for polynomials imply that the analytic complexity of~$p_0(x,y)$ does not exceed~$5.$ Indeed, the support of~$p_0(x,y)$ lies in the union of~$10$ lines parallel to $x$ axis. The analytic complexity of polynomial whose support lies on a straight line parallel to axis cannot exceed~$1.$ Then the analytic complexity of the sum of $k$ such polynomials cannot exceed~$1+\lceil \log_2 k\rceil.$  Later we prove that the analytic complexity of~$p_0(x,y)$ is actually equal to~$3.$ 
 \label{ex:SimpleZonotopeSystem}
\end{example}

In general, appending a pair of rows~$(a_i, b_i),(-a_i,-b_i)$ to the matrix defining a hypergeometric system is equivalent to adding a pair of parallel divisors in the exponent space. Let the hypergeometric system $\text{Horn}(A_0,c_0)$ be defined by a parallelogram, and $p_0(x,y) = \sum\limits_{(s,t)\in S}c_{s,t}\cdot x^sy^t$ be a polynomial solution of this system, $S$ be its support. Adding a pair of divisors in the exponent space leads to the system with the solution given by

$$p_1(x,y)=\sum\limits_{(s,t)\in S}\frac{{\rm\Gamma}\left(\alpha_1 s+\beta_1 t  + \gamma_1 + 1\right)}{{\rm\Gamma}\left(\alpha_1 s+\beta_1 t  + \gamma_1\right)}\cdot c_{s,t}\cdot x^s y^t = \sum\limits_{(s,t)\in S} \left(\alpha_1 s + \beta_1 t + \gamma_1\right) x^s y^t$$

$$=\left(\alpha_1 \theta_x + \beta_1 \theta_y + \gamma_1\right)\sum\limits_{(s,t)\in S}  c_{s,t} x^s y^t = \left(\alpha_1 \theta_x + \beta_1 \theta_y + \gamma_1\right) p_0(x,y).$$

Using this formula repetitively we obtain the solution for $k$~pairs of additional divisors:

$$p_k(x,y) = \left(\prod\limits_{j=1}^{k}\left(\alpha_j \theta_x+\beta_j\theta_y+\gamma_j\right)\right) p_0(x,y).$$

Thus the estimate for the analytic complexity of $p_k(x,y)$ depends on the analytic complexity of $p_0(x,y).$ This dependence is described in detail in the following Proposition and its corollaries. There $\theta_x=x\frac{\partial}{\partial x}, \theta_y=y\frac{\partial}{\partial y}$ and $\alpha_i,\beta_i,\gamma_i\in\mathbb{C}, i=0,1,\ldots$

\begin{proposition} \rm
Let $f(x,y)$ be a $Cl_n$ function. Then $$(\alpha_0 \theta_x + \beta_0 \theta_y + \gamma_0) f(x,y) \in Cl_{2n+1}.$$
{\it Proof.} The proof of the statement is based on the proof of Proposition 8 in~\cite{BeloshapkaRJMP2012}. Consider the result of the differential operator $\alpha_0\theta_x+\beta_0\theta_y$ action on the function $f(x,y).$ Using the induction by $n$ we can prove that this function belongs to~$Cl_{2n}.$

 For $n=1$ we can represent $f(x,y)$ in the form $f(x,y)=c(a(x)+b(y)).$ $$(\alpha_0\theta_x+\beta_0\theta_y) c(a(x)+b(y)) = c'(a(x)+b(y))\cdot\left(\alpha_0 xa'(x)+\beta_0 yb'(y)\right),$$ and this function belongs to $Cl_2$ as a product of $Cl_1$ functions. If the statement holds for all $n<N,$ and $f(x,y)$ belongs to $Cl_N,$ which means it can be represented as $f(x,y)=h(f_1(x,y)+f_2(x,y)),$ where $f_1(x,y),f_2(x,y)\in Cl_{N-1},$ then $$(\alpha_0\theta_x+\beta_0\theta_y) h(f_1(x,y)+f_2(x,y))=h'(f_1(x,y)+f_2(x,y))\cdot $$ $$\cdot\left((\alpha_0 \theta_x + \beta_0 \theta_y) f_1(x,y) + (\alpha_0 \theta_x + \beta_0 \theta_y) f_2(x,y) \right).$$ Both of the functions $f_1(x,y)$ and $f_2(x,y)$ belong to $Cl_{N-1},$ so the estimate of the analytic complexity for  $(\alpha_0 \theta_x + \beta_0 \theta_y) f_i(x,y),i=1,2$ is $Cl_{2N-2}.$ Then their sum belongs to $Cl_{2N-1}$ and, after the multiplication of the result by $h'(f_1(x,y)+f_2(x,y))\in Cl_N,$ the product belongs to $Cl_{2N}.$ Thus we conclude that for any $n,$ if $f(x,y)\in Cl_n$ then $(\alpha_0 \theta_x + \beta_0 \theta_y) f(x,y)\in Cl_{2n}.$  Adding $\gamma_0 f(x,y) \in Cl_n$ to this expression we obtain a~function in~$Cl_{2n+1}.$~\hfill $\square$
\end{proposition}

\begin{corollary} \rm
For any $f(x,y)\in Cl_n$ the analytic complexity of $$\left(\prod\limits_{j=1}^{k}\left(\alpha_j \theta_x+\beta_j\theta_y+\gamma_j\right)\right)f(x,y)$$ cannot exceed $2^k(n+1)-1.$
\end{corollary}

\begin{corollary} \rm
Assume that the analytic complexity of a polynomial solution $p_0(x,y)$ to the hypergeometric system~$\text{Horn}(A_0,c_0)$ does not exceed $n,\; S$ is a support of~$p_0(x,y).$ Let the matrix~$A$ be obtained from~$A_0$ by appending~$k$ pairs of vectors $(a_i,b_i),(-a_i,-b_i),$ vector~$c$ be obtained from~$c_0$ by appending~$2k$ elements. Then the analytic complexity of a polynomial solution with the support~$S$ to the hypergeometric system~$\text{Horn}(A,c)$ does not exceed~$2^k(n+1)-1.$
\label{cor:AnalyticComplexityEstimate}
\end{corollary}

While this estimate is rough when we use it for several additional pairs of divisors ($k > 1$), for $k=1$ it can be quite accurate.

\vskip.3cm
{\it\noindent Example~\ref{ex:SimpleZonotopeSystem}. (Continued).}
 Let us use Corollary~\ref{cor:AnalyticComplexityEstimate} to estimate the analytic complexity of a solution to the system $\text{Horn}(A_0,c_0).$ To do this, consider the system $\text{Horn}(\tilde{A}_0,\tilde{c}_0),$ defined by the matrix $\tilde{A}_0= \left(\begin{array}{rrrr} 1 & 0 & 0 & -1 \\ 0 & 1 & -1 & 0 \end{array}\right)^{\rm T}$ and the vector of parameters $\tilde{c}_0=(-10,-9, 1, 1).$ This system differs from the original one only by an absence of the pair of divisors with the normal vectors $(1,1)$ and $(-1,-1).$ Thus the support of the solution to the system $\text{Horn}(\tilde{A}_0,\tilde{c}_0)$ coincides with the support of~$p_0(x,y).$  Note that this system is defined by a parallelogram and hence by Proposition~\ref{prop:parallelogram} the analytic complexity of its solutions cannot exceed~$2.$ Computations show that the basis in the space of solutions to the system~$\text{Horn}(\tilde{A}_0,\tilde{c}_0)$ consists only of one function: $(x-1)^{10}(y-1)^9 \in Cl_1,$ then $p_0(x,y)\in Cl_3$ by Corollary~\ref{cor:AnalyticComplexityEstimate}. Supports of two other solutions to~$\text{Horn}(A_0,c_0)$ lie on two parallel divisors, so a linear combination of these solutions belongs to~$Cl_3,$ and the general solution to~$\text{Horn}(A_0,c_0)$ is a function in~$Cl_4.$

The resulting analytic complexity estimate of solutions to hypergeometric systems defined by zonotopes is formulated in the following theorem.

\begin{theorem} \rm
Let $\text{Horn}(A,c)$ be a hypergeometric system and~$\text{Horn}(A,c)$ is defined by a zonotope. Assuming the matrix~$A$ contains~$2k$ rows, consider matrices~$\hat{A}$ and $-\hat{A}$ such that the union of their rows coincides with the set of rows of~$A.$ Let $\alpha$ be a part of the parameter vector~$c,$ corresponding to the matrix $\hat{A},\; \beta$ be a part of this vector, corresponding to~$-\hat{A},$ and define the vector $\hat{c}$ with elements $\hat{c}_i=-\alpha_i-\beta_i.$ 

If $\hat{c}_i \in \mathbb{N}\backslash\{0\},i=1,\ldots,k,$ then the analytic complexity of the general solution to~$\text{Horn}(A,c)$ does not exceed $$\min\left(3\cdot 2^{k-2}-1+\lceil \log_2\frac{k(k-1)}{2}\rceil,2+\lceil\log_2(\max\limits_i\hat{c}_i+1)\rceil+\lceil\log_2(k-1)\rceil\right).$$

\noindent{\it Proof.} The condition $\hat{c}_i \in \mathbb{N}\backslash\{0\}$ provides the existence of a polynomial basis in the space of solutions to~$\text{Horn}(A,c).$ The matrix~$A$ contains $2k$ rows, so supports of the solutions are bounded by $k$~pairs of divisors. The union of these supports is a subset of~$\frac{k(k-1)}{2}$ parallelogram intersections (it is a sum of an arithmetic progression) and in every intersection the solution belongs to~$Cl_{3\cdot 2^{k-2}-1}$ (by Corollary~\ref{cor:AnalyticComplexityEstimate}). 

On the other hand, there is the estimate based on the number of parallel lines connecting the points of the support (see Proposition~4 in~\cite{BeloshapkaRJMP2012}). While the analytic complexity of any polynomial with the support belonging to a straight line does not exceed~$2,$ the number of these lines for every pair of divisors equals~$\hat{c}_i+1.$ Thus for any pair of divisors, the part of the solution, belonging to intersections of this pair and any other pairs cannot exceed~$2+\lceil\log_2(\max\limits_i\hat{c}_i+1)\rceil.$ Note that there is no need to use all of~$k$ pairs of divisors to estimate the analytic complexity of the general solution this way, since $k-1$ pairs already bound the whole support of the solution.

To obtain the estimate from the statement of Theorem, we have to find the estimate based on pairing of $\frac{k(k-1)}{2}$ parallelogram intersections, then to find the estimate, based on pairing of $k-1$ pairs of the divisors. Minimal of those numbers is the sought estimate.~\hfill $\square$
\label{thm:ZonotopeAnalyticComplexity}
\end{theorem}

Let us order $\hat{c}_i$ by the ascension and then $v$ be a vector with the elements $v_i=\min\left(2+\lceil\log_2(\hat{c}_i+1)\rceil, 3\cdot 2^{k-2}-1+\right.$ $\left.\lceil \log_2(k-i)\rceil \right),i=1,\ldots,k-1.$ To find more accurate value for the analytic complexity estimate from Theorem~\ref{thm:ZonotopeAnalyticComplexity}, one could use Algorithm~\ref{alg:SumAnalyticComplexity} from Section~\ref{sec:Algorithms} using $v$ as an input vector. The accuracy is obtained due to the fact that the vector $v$ provides the decision of better estimate for every pair of divisors, since $\hat{c}_i$ may have high values not for all of them. 


\section{Algorithms of Analytic Complexity Estimation}
\label{sec:Algorithms}

To estimate the analytic complexity of the general solution to the hypergeometric system from Theorem~\ref{thm:ZonotopeAnalyticComplexity} one can use the following algorithm. 

\begin{algorithm}[H]
\DontPrintSemicolon
  
  \KwInput{$c=\{c_1,c_2,\ldots,c_n\}$ - a set of known estimates of the analytic complexity values for bivariate functions~$f_1(x,y),f_2(x,y),\ldots,f_n(x,y),$ where $(x, y)\in\mathbb{C}^2.$}
  \KwOutput{$N$ - an estimate for the analytic complexity of the function $\sum\limits_{i=1}^n f_i(x,y).$}

  \While{$c$ {\rm contains more than}~$1$ {\rm element}}
  {
  
  find $2$ minimal elements of $c,$ namely, $c_i$ and $c_j.$
  
  $c = (c \cup \{\max(c_i,c_j)+1\}) \backslash  \{c_i, c_j\}.$ 
  
  }
  
  $N \leftarrow$ only element of~$c.$
    
\caption{Finding the analytic complexity estimate for the sum of functions}
\label{alg:SumAnalyticComplexity}
\end{algorithm}
Algorithm~\ref{alg:SumAnalyticComplexity} is finite, since at each step the number of elements in~$c$ decreases~by~$1.$

The following algorithm allows one to find the support of a polynomial solution to a given hypergeometric system defined by a zonotope, provided that such a solution exists. The algorithm is based on Proposition~4.7 in~\cite{Sadykov-Tanabe}. 

\begin{algorithm}[H]
\DontPrintSemicolon
  
  \KwInput{ the matrix $A,$ the parameter vector~$c$ for the hypergeometric system $\text{Horn}(A,c)$ defined by a zonotope}
  \KwOutput{$supp$ - the support for the polynomial solution to~$\text{Horn}(A,c).$}
  
  $supp \leftarrow \{\}$
  
  find $\hat{A}:$ rows$(\hat{A})\cup$ rows$(-\hat{A})$ = rows$(A)$
  
  \For {$(r_i, r_j) \subset\text{\rm rows} (\hat{A}), i < j$}
  {
  
  $A_{i,j} \leftarrow (r_i,r_j)^T$

  $\alpha \leftarrow$ elements of $c$ corresponding to~$(r_i, r_j)$
  
  $\beta \leftarrow$ elements of $c$ corresponding to~$(-r_i,-r_j)$
 
  \If{$-\alpha_j-\beta_j > 0 \text{ for } j=1,2$}{ 
  $supp = supp \cup \text{Supp} \left( x^{-A_{i,j}^{-1}\alpha}\left(1+x^{-A_{i,j}^{-1}e_1}\right)^{-\alpha_1-\beta_1}\left(1+x^{-A_{i,j}^{-1}e_2}\right)^{-\alpha_2-\beta_2}\right)$}
  \Else{ the general solution to $\text{Horn}(A,c)$ is not a polynomial}	

  }
\caption{Constructing the support for the polynomial solution to the hypergeometric system}
\label{alg:SupportForPolynomialSolution}
\end{algorithm}

For some pairs of rows $r_i, r_j$ solution to the corresponding system defined by a parallelogram is not a polynomial. In this case, part of the basis in the solution space can still consists of polynomials, and their supports can be found by the means of Algorithm~\ref{alg:SupportForPolynomialSolution}.

The following algorithm allows one to compute the analytic complexity of any given bivariate polynomial. 

\begin{algorithm}[H]
\DontPrintSemicolon
  
  \KwInput{$p(x,y)$ - a polynomial, $x, y\in\mathbb{C}.$}
  \KwOutput{$N$ - an estimate for the analytic complexity of $p(x,y).$}
  $result \leftarrow 0$
  
  $short \leftarrow \{\}$
  
  $polys \leftarrow  \{p_i(x,y)| p(x,y)=\sum\limits_i p_i(x,y), \text{Supp }p_i(x,y)||\text{Supp }p_j(x,y) \forall i,j \} $

   \For{$p \in polys$}    
   { 
      	$curr = getShort(p)$
      	
      	\If{$curr \not\subset short$}
      	{
		result += 1	      		
      	}
      	
      	$short = short \cup curr$
      	
   }
   
   $N \leftarrow 2 + \lceil Log_2(result)\rceil$

\caption{Finding the analytic complexity estimate for the polynomial}
\end{algorithm}

The main improvement of this algorithm compared to the existing ones is its ability to distinct the powers of lower degree polynomials included in the original polynomial as summands. Without this feature, even the analytic complexity of the function like~$p(a(x)+b(y))\in Cl_1,$ where $p(t),a(x),b(y)$ are univariate polynomials, is estimated based on its support, which becomes very complex with the growth of degree of~$p(t).$    

The input of the function $getShort()$ is a homogeneous polynomial and the output contains elements of its decomposition into the sum of powers. Note that the definition of $polys$ assumes the ambiguity of the representation of the polynomial as the sum of finitely many polynomials with their supports lying in parallel straight lines. Any of such representations give an estimate, but some of them may be better than other ones.


%
%

\section{Examples}
\label{sec:Examples}

{\it\noindent Example~\ref{ex:SimpleZonotopeSystem}. (Continuation).} Let us replace the parameter vector~$c_0$ in the system~$\text{Horn}(A_0,c_0)$ by the vector~$(k,0,0,0,0,0).$ The corresponding system is given~by
$$
\begin{array}{l}
x \theta_{x} (\theta_x + \theta_y + k) - \theta_{x} (\theta_x + \theta_y), \\
y \theta_{y} (\theta_x + \theta_y + k) - \theta_{y} (\theta_x + \theta_y).
\end{array}
$$
A basis in its solution space is given by $1,$ $ \log\frac{x}{x-1} + \sum_{j=1}^{k-1} \frac{(-1)^j}{j(x-1)^j},$
$ \log\frac{y}{y-1} + \sum_{j=1}^{k-1} \frac{(-1)^j}{j(y-1)^j},$ so there is no polynomial basis for these parameter values. Nevertheless, the analytic complexity of the general solution is equal to~$1.$ 

The present example shows that the analytic complexity of solutions to hypergeometric systems can be heavily dependent on parameter vectors defining these systems. A resonant choice of their parameters can drastically reduce the analytic complexity of general solutions to such systems.

\begin{example} {\it An octagon zonotope.}
Consider Example 6.8 in~\cite{Sadykov-Tanabe}. In order to find the analytic complexity of a polynomial solution to the hypergeometric system defined by the matrix 
$$A=\left(\begin{array}{rrrrrrrr}1 & -1 & -1 & 1 & -3 & 3 & 2 & -2 \\ 2 & -2 & 1 & -1 & -2 & 2 & -1 & 1 \end{array}\right)^{\rm T}$$ and the vector of parameters $c=(3,-5,-2,1,-2,-1,-1,-1)$ we can use the basis of the solutions to this system, computed in the book. There are $3$~solutions whose analytic complexity equal to~$2$, and $28$~solutions in~$Cl_1,$ two of them also belonging to $Cl_0.$ Therefore the analytic complexity of the general solution to this system cannot exceed~$7.$ Note that this estimate is based on a~trivial pairing of the basis functions, but very specific structure of the solution support makes it possible to estimate the analytic complexity not to exceed~$6.$


Let us estimate the analytic complexity of the general solution to this system, using Theorem~\ref{thm:ZonotopeAnalyticComplexity}. The vector $\hat{c},$ ordered by the ascension, is $(1,2,2,3).$ Then the vector~$v=(3,4,4)$ (it includes only support-based estimates, because of low values of the elements of~$\hat{c}),$ and, by using Algorithm~\ref{alg:SumAnalyticComplexity}, we conclude that the general solution belongs to~$Cl_6.$ Note that this estimate coincides with the one we have obtained by hand.

Futhermore, we can estimate the analytic complexity of a solution to any hypergeometric system we obtain by appending a pair of rows to~$A$ (the only condition is that these rows are not collinear to the rows of~$A$). Note that this estimate does not depend much on the difference between new parameters. If this difference is great, it becomes the last element of the ordered vector $\hat{c},$ and does not affect the new vector~$v,$ the new element of the vector~$v$ is equal to~$2+\lceil\log_2(3+1)\rceil=4,$ and the resulting analytic complexity is $6.$ On the contrary, if this difference is low, for example, if it is equal to~$1,$ the new vector~$\hat{c}=(1,1,2,2,3),$ the new vector~$v=(3,3,4,4),$ and the analytic complexity is also equal to~$6.$ Thus we conclude that the addition of $2$ rows to the matrix~$A$ does not affect the analytic complexity of the solution to the system.
\end{example}

\begin{example}  {\it A decagon zonotope.}
Consider the hypergeometric system~$\text{Horn}(A_1,c_1),$ defined by the matrix \begin{equation} A_1=\left(\begin{array}{rrrrrrrrrr} -1 & 1 & 0 & 0 & -2 & 2 & 3 & -3 & 3& -3\\ 0 & 0 & -1 & 1 & 1 & -1 & 1 & -1 & 2 & -2 \end{array}\right)^{\rm T} \label{eq:A1Matrix} \end{equation} and the parameter vector~$c_1 = (-1,0,4,-5,1,-4,-9,6,-4,0).$ The zonotope defining the matrix~$A_1$ is shown in Figure~\ref{fig:zonotopeComplexExample}. 

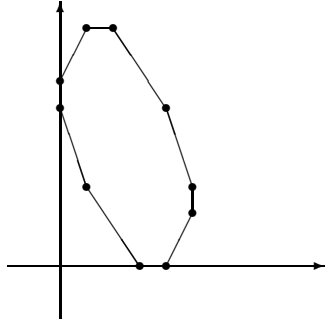
\begin{figure}[htbp]
\centering
\begin{minipage}{5cm}
\begin{picture}(120,120)
  \put(20,0){\vector(0,1){120}}
  \put(0,20){\vector(1,0){120}}
 


  \put(60,20){\line(1,2){10}}
  \put(70,40){\line(0,1){10}}
  \put(70,50){\line(-1,3){10}}
  \put(60,80){\line(-2,3){20}}
  \put(40,110){\line(-1,0){10}}
  \put(30,110){\line(-1,-2){10}}
  \put(20,80){\line(1,-3){10}}
  \put(30,50){\line(2,-3){20}}
  \put(50,20){\circle*{3}}
  \put(60,20){\circle*{3}}
  \put(70,40){\circle*{3}}
  \put(70,50){\circle*{3}}
  \put(60,80){\circle*{3}} 
  \put(40,110){\circle*{3}} 
  \put(30,110){\circle*{3}}
  \put(20,90){\circle*{3}} 
  \put(20,80){\circle*{3}}  
  \put(30,50){\circle*{3}}
\end{picture}
\end{minipage}
\caption{The zonotope which defines the matrix~(\ref{eq:A1Matrix})}
\label{fig:zonotopeComplexExample}
\end{figure}

The holonomic rank of the system~$\text{Horn}(A_1,c_1)$ equals~$34.$ The support to the solution of this system computed by the means of Algorithm~\ref{alg:SupportForPolynomialSolution} is shown in Figure~\ref{fig:supportComplexExample}.

\begin{figure}[htbp]
\centering
\begin{minipage}{8cm}
\begin{picture}(250,250)
  \put(125,0){\vector(0,1){260}}
  \put(0,125){\vector(1,0){260}}
  \put(119,117){\small $0$}
  \put(5,125){\line(0,1){3}}
  \put(25,125){\line(0,1){5}}
  \put(20,116){\small $-5$}
  \put(45,125){\line(0,1){3}}
  \put(65,125){\line(0,1){3}}
  \put(85,125){\line(0,1){3}}
  \put(105,125){\line(0,1){3}}
  \put(145,125){\line(0,1){3}}
  \put(165,125){\line(0,1){3}}
  \put(185,125){\line(0,1){3}}
  \put(205,125){\line(0,1){3}}
  \put(225,125){\line(0,1){5}}
  \put(222,116){\small $5$}
  \put(245,125){\line(0,1){3}}
  \put(125,5){\line(1,0){3}}
  \put(125,15){\line(1,0){3}}
  \put(125,25){\line(1,0){5}}
  \put(106,21){\small $-10$}
  \put(125,35){\line(1,0){3}}  
  \put(125,45){\line(1,0){3}}
  \put(125,55){\line(1,0){3}}
  \put(125,65){\line(1,0){3}} 
  \put(125,75){\line(1,0){5}} 
  \put(125,85){\line(1,0){3}}
  \put(125,95){\line(1,0){3}}
  \put(125,105){\line(1,0){3}}
  \put(125,115){\line(1,0){3}}
  \put(125,135){\line(1,0){3}} 
  \put(125,145){\line(1,0){3}}
  \put(125,155){\line(1,0){3}}
  \put(125,165){\line(1,0){3}} 
  \put(125,175){\line(1,0){5}} 
  \put(125,185){\line(1,0){3}}
  \put(125,195){\line(1,0){3}}
  \put(125,205){\line(1,0){3}} 
  \put(125,215){\line(1,0){3}} 
  \put(125,225){\line(1,0){5}}
  \put(132,222){\small $10$}
  \put(125,235){\line(1,0){3}}
  \put(125,245){\line(1,0){3}}  
  \put(105,0){\line(0,1){260}}
  \put(0,165){\line(1,0){260}}
  \put(0,175){\line(1,0){260}}
  \put(10,0){\line(1,1){250}}
  \put(20,0){\line(1,1){240}}
  \put(30,0){\line(1,1){230}}
  \put(40,0){\line(1,1){220}}
  \put(0,239){\line(4,-3){260}}
  \put(0,234){\line(4,-3){260}}
  \put(0,229){\line(4,-3){260}}  
  \put(0,224){\line(4,-3){260}}
  \put(0,219){\line(4,-3){260}}  
  \put(95,260){\line(2,-3){165}}
  \put(88,260){\line(2,-3){174}}
  \put(81,260){\line(2,-3){174}}
  \put(74,260){\line(2,-3){174}}
  \put(86,175){\circle*{3}}
  \put(79,175){\circle*{3}}
  \put(72,175){\circle*{3}}
  \put(65,175){\circle*{3}}
  \put(58,175){\circle*{3}}
  \put(100,165){\circle*{3}}
  \put(93,165){\circle*{3}}
  \put(86,165){\circle*{3}}
  \put(79,165){\circle*{3}}
  \put(72,165){\circle*{3}}  
  \put(131,175){\circle*{3}}
  \put(138,175){\circle*{3}}
  \put(145,175){\circle*{3}}
  \put(152,175){\circle*{3}}
  \put(138,165){\circle*{3}}  
  \put(145,165){\circle*{3}}
  \put(152,165){\circle*{3}}
  \put(159,165){\circle*{3}}  
  \put(205,165){\circle*{3}}
  \put(195,165){\circle*{3}}
  \put(185,165){\circle*{3}}
  \put(175,165){\circle*{3}}
  \put(185,175){\circle*{3}}
  \put(195,175){\circle*{3}}
  \put(205,175){\circle*{3}}
  \put(215,175){\circle*{3}}
  \put(131,121){\circle*{3}}
  \put(134,124){\circle*{3}}
  \put(137,127){\circle*{3}}
  \put(137,117){\circle*{3}}
  \put(140,130){\circle*{3}}
  \put(140,120){\circle*{3}}
  \put(143,133){\circle*{3}}
  \put(143,123){\circle*{3}}
  \put(142,112){\circle*{3}}
  \put(146,126){\circle*{3}}
  \put(145,115){\circle*{3}}     
  \put(149,129){\circle*{3}}
  \put(148,118){\circle*{3}}
  \put(151,121){\circle*{3}}
  \put(154,124){\circle*{3}}
  \put(148,108){\circle*{3}}
  \put(151,111){\circle*{3}}
  \put(154,114){\circle*{3}}
  \put(157,117){\circle*{3}}
  \put(160,120){\circle*{3}}

  \put(165,125){\circle*{3}}
  \put(153,142){\circle*{3}}
  \put(157,146){\circle*{3}}
  \put(157,136){\circle*{3}}
  \put(161,150){\circle*{3}}
  \put(161,140){\circle*{3}}
  \put(161,130){\circle*{3}}
  \put(165,155){\circle*{3}}
  \put(165,145){\circle*{3}}
  \put(165,135){\circle*{3}}
  \put(169,149){\circle*{3}}
  \put(169,139){\circle*{3}}
  \put(169,129){\circle*{3}}
  \put(173,143){\circle*{3}}
  \put(173,133){\circle*{3}}
  \put(177,137){\circle*{3}}
  \put(125,215){\circle*{3}}
  \put(125,205){\circle*{3}}
  \put(125,195){\circle*{3}}
  \put(125,185){\circle*{3}}
  \put(105,245){\circle*{3}}
  \put(105,235){\circle*{3}}
  \put(105,225){\circle*{3}}
  \put(105,215){\circle*{3}}
  \put(125,165){\circle*{3}}
  \put(125,175){\circle*{3}}
  \put(105,165){\circle*{3}}
  \put(105,175){\circle*{3}}
  \put(105,160){\circle*{3}}
  \put(105,155){\circle*{3}}
  \put(105,150){\circle*{3}}
  \put(105,145){\circle*{3}}
  \put(105,140){\circle*{3}}
  \put(125,145){\circle*{3}}
  \put(125,140){\circle*{3}}
  \put(125,135){\circle*{3}}
  \put(125,130){\circle*{3}}
  \put(125,125){\circle*{3}}
  \put(125,115){\circle*{3}}
  \put(125,105){\circle*{3}}
  \put(125,95){\circle*{3}}
  \put(125,85){\circle*{3}}
  \put(105,95){\circle*{3}}
  \put(105,85){\circle*{3}}
  \put(105,75){\circle*{3}}
  \put(105,65){\circle*{3}}

  \put(176,108){\circle*{3}}
  \put(183,97){\circle*{3}}
  \put(189,87){\circle*{3}} 
  \put(196,77){\circle*{3}} 
  \put(203,66){\circle*{3}}  
  \put(190,97){\circle*{3}}
  \put(196,87){\circle*{3}} 
  \put(203,77){\circle*{3}} 
  \put(210,66){\circle*{3}}
  \put(217,56){\circle*{3}}
  \put(203,87){\circle*{3}} 
  \put(210,77){\circle*{3}} 
  \put(217,66){\circle*{3}}
  \put(224,56){\circle*{3}}
  \put(231,45){\circle*{3}}
  \put(217,77){\circle*{3}} 
  \put(224,66){\circle*{3}}
  \put(231,56){\circle*{3}}
  \put(238,45){\circle*{3}}
  \put(245,35){\circle*{3}}
\end{picture}
\end{minipage}
\caption{The support for the solution of the system $\text{Horn}(A_1,c_1)$}
\label{fig:supportComplexExample}
\end{figure}
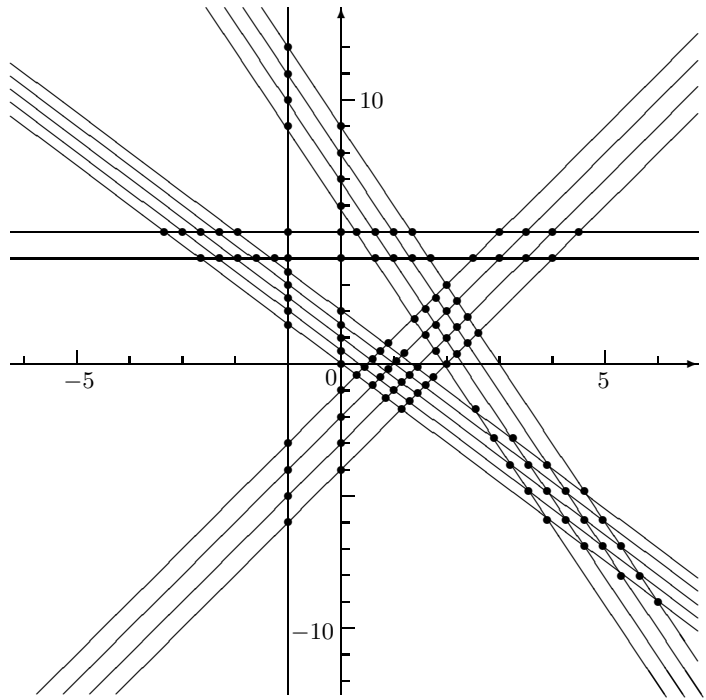

Polynomial basis in the solution space to~$\text{Horn}(A_1,c_1)$ consists of the $4$~monomials $\frac{x^6}{y^9},\frac{x^{17/3}}{y^8},\frac{x^3}{y^3},\frac{x^{8/3}}{y^2}$ and $30$~polynomials $$\frac{1}{x y^6}+\frac{5643}{637 x y^5}+\frac{247095}{8281 x y^4}+\frac{329460}{8281 x y^3}+\frac{27455}{286 y^4}+\frac{82365}{49 y^3}+\frac{741285}{49 y^2}+\frac{724812}{7 y},$$ $$\frac{20 y^3}{63 x}-\frac{4 y^2}{35 x}+\frac{4 y^2}{5}-\frac{18 y}{5}+1,\frac{3 y^{3/2}}{380 x}-\frac{3969 y^{5/2}}{41990 x}+\frac{1323 y^{7/2}}{16796 x}-\frac{51}{55} y^{3/2}+ \sqrt{y},$$ $$-\frac{11 y^{12}}{115311 x}-\frac{33 y^{11}}{38437 x}-\frac{297 y^{10}}{100555 x}-\frac{24 y^9}{5915 x}+\frac{3 y^9}{1105}+\frac{y^8}{26}+\frac{36 y^7}{143}+y^6, x y^4-\frac{2}{13} x y^5,$$ $$\frac{8 y^5}{99 x}+\frac{4 y^4}{3 x}+\frac{50y^5}{81}+ y^4, \frac{1550775 x^{7/2} y^5}{82808479}-\frac{31465  x^{9/2} y^5}{61400001}+ x^{5/2} y^4-\frac{5175 x^{7/2} y^4}{89947},$$ $$\frac{1547 x^4 y^5}{103455}-\frac{91 x^4 y^4}{6840}-\frac{91 x^3 y^5}{1026}+ x^3 y^4, \frac{806 y^5}{129 x^{8/3}}+\frac{84656 y^4}{735 x^{5/3}}+\frac{y^4}{x^{8/3}},\frac{x^{13/3}}{y^6}+\frac{451 x^{13/3}}{261 y^5},$$ $$\frac{87 y^5}{82 x^{7/3}}+\frac{5220 y^5}{275561 x^{10/3}}+\frac{36575 y^4}{2392 x^{4/3}}+\frac{y^4}{x^{7/3}},\frac{44 y^5}{1183 x^3}+\frac{33 y^5}{182 x^2}+\frac{y^4}{x^2},\frac{x^{16/3}}{y^8}+\frac{1378 x^{16/3}}{451 y^7},$$ $$-\frac{21}{46} x^{2/3} y^5+x^{2/3} y^4+\frac{119}{286} x^{5/3} y^4,-\frac{12}{247} x^{4/3} y^5+x^{4/3} y^4-\frac{364 \sqrt[3]{x} y^5}{1045},\frac{2 x}{7 y}+ x,$$ $$\frac{11985}{299} x^{8/7} y^{2/7}+\frac{14382 x^{8/7}}{253 y^{5/7}}+\frac{x^{8/7}}{y^{12/7}},\frac{1200 x^{2/7}}{1643 y^{3/7}}+\frac{345 x^{9/7}}{31 y^{3/7}}+\frac{x^{9/7}}{y^{10/7}},\frac{x^{10/3}}{y^4}+\frac{261 x^{10/3}}{238 y^3},$$ $$\frac{114774 x^{6/7} y^{5/7}}{28405}+\frac{1188 x^{6/7}}{65 y^{2/7}}+\frac{x^{6/7}}{y^{9/7}},\frac{x^{10/7}}{y^{8/7}}+\frac{731 x^{3/7}}{638 \sqrt[7]{y}}+\frac{1763 x^{10/7}}{754 \sqrt[7]{y}},$$ $$\frac{x^{4/7}}{y^{6/7}}+\frac{32680 x^{11/7}}{8613 y^{6/7}}+\frac{1558}{261} x^{4/7} \sqrt[7]{y},\frac{169}{150} x^{5/7} y^{3/7}+\frac{x^{5/7}}{y^{4/7}}+\frac{65 x^{12/7}}{136 y^{4/7}},$$ $$-\frac{1}{66} 5 x^2 y^3+\frac{5}{7} x^2 y^2-\frac{45}{28} x^2 y+ x^2,x^{11/5} y^{2/5}-\frac{4301 x^{11/5} y^{7/5}}{4277}+\frac{232254 x^{11/5} y^{12/5}}{1056419},$$ $$x^{9/5} y^{3/5}-\frac{1287 x^{9/5} y^{8/5}}{1634}+\frac{55913 x^{9/5} y^{13/5}}{346408},x^{12/5} y^{4/5}-\frac{68}{19} x^{7/5} y^{9/5}-\frac{116}{231} x^{12/5} y^{9/5},$$ $$x^{8/5} y^{6/5}+\frac{5824 x^{13/5} y^{6/5}}{432837}-\frac{1064 x^{8/5} y^{11/5}}{2829},\frac{x^5}{y^7}+\frac{8 x^5}{15 y^6}-\frac{21 x^4}{55 y^6}-\frac{182 x^4}{15 y^5}-\frac{91 x^4}{24 y^4},$$ $$\frac{x^{14/3}}{y^7}+\frac{828 x^{14/3}}{85 y^6}-\frac{585488 x^{11/3}}{48825 y^5}+\frac{21758 x^{14/3}}{23715 y^5}-\frac{2488324 x^{11/3}}{35805 y^4}.$$ There are $14$ functions in~$Cl_1$ and $20$~functions in~$Cl_2\backslash Cl_1$ among these polynomials. 

The analytic complexity estimate of the general solution obtained by the pairing of these functions is~$Cl_7.$ Theorem~\ref{thm:ZonotopeAnalyticComplexity} gives the following estimate: $\hat{c} = (2,2,3,3,4), v=(4,4,4,4),$ then the general solution belongs to~$Cl_6.$

\end{example}

The following examples present hypergeometric systems defined by non-zonotope polygons, with solutions having low analytic complexity.

\begin{example} {\it A pentagon.}
\rm The matrix $\left(\begin{array}{rrrrrrrrrr} 1 & -1 & 0 & 1 & -1 & 0 & 0\\ 1 & 0 & -1 & 0 & 0 & -1 & 1 \end{array}\right)^{\rm T}$ and the vector of parameters $(-4,0,0,-1,-2,-1,-2)$ define the hypergeometric system
\begin{equation}
\begin{array}{l}
x (\theta_x + \theta_y -4) (\theta_x - 1) - \theta_x (\theta_x - 2), \\
y (\theta_x + \theta_y -4) (\theta_y - 1) - \theta_y (\theta_y - 2).
\end{array}
\label{eq:NonZonotopeSystem}
\end{equation}
This system is holonomic and its holonomic rank equals~4. The pure basis (see~\cite{Sadykov-Tanabe}) in its
solution space is given by the Taylor polynomials
$$
x^2 y^2, \; 1 - 4 x - 4 y + 12 x y, \; 6x^2 - 4x^3 + x^4 - 12x^2 y + 4 x^3 y, \;
6y^2 - 12x y^2 - 4y^3 + 4x y^3 + y^4.
$$ The first and the second of these polynomials belong to~$Cl_1,$ the third and the fourth belong to~$Cl_2.$ Thus the general solution is a function in~$Cl_4.$ 
\begin{figure}
\centering
\begin{minipage}{4cm}
\begin{picture}(60,60)
  \put(10,0){\vector(0,1){60}}
  \put(0,10){\vector(1,0){60}}
  \put(54,12){\small $s$}
  \put(12,54){\small $t$}
  \put(0,60){\line(1,-1){60}}
  \put(20,0){\line(0,1){50}}
  \put(30,00){\line(0,1){40}}
  \put(0,20){\line(1,0){50}}
  \put(0,30){\line(1,0){40}}
  \put(10,10){\circle*{3}}
  \put(10,20){\circle*{3}}
  \put(20,10){\circle*{3}}
  \put(20,20){\circle*{3}}
  \put(30,10){\circle{5}}
  \put(30,20){\circle{5}}
  \put(40,10){\circle{5}}
  \put(40,20){\circle{5}}
  \put(50,10){\circle{5}}
  \put(10,30){\circle*{5}}
  \put(20,30){\circle*{5}}
  \put(10,40){\circle*{5}}
  \put(20,40){\circle*{5}}
  \put(10,50){\circle*{5}}
  \put(30,30){\circle*{7}}
\end{picture}
\end{minipage}
\hskip3cm
\begin{minipage}{3cm}
\begin{picture}(40,40)
  \put(10,0){\vector(0,1){40}}
  \put(0,10){\vector(1,0){40}}
  \put(30,10){\line(0,1){10}}
  \put(30,20){\line(-1,1){10}}
  \put(20,30){\line(-1,0){10}}
  \put(10,10){\circle*{3}}
  \put(10,20){\circle*{3}}
  \put(10,30){\circle*{3}}
  \put(20,10){\circle*{3}}
  \put(20,20){\circle*{3}}
  \put(20,30){\circle*{3}}
  \put(30,10){\circle*{3}}
  \put(30,20){\circle*{3}}
\end{picture}
\end{minipage}
\caption{a): the supports of solutions to the system~(\ref{eq:NonZonotopeSystem}); b) polygon defining the system~(\ref{eq:NonZonotopeSystem})}
\end{figure}
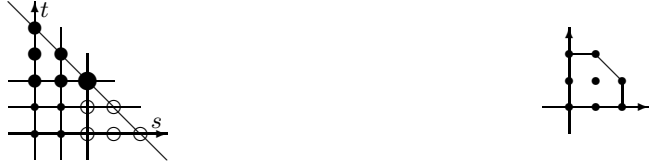
\label{ex(1,1)(1,0),(0,1),(-1,0)^2,(0,1)^2}
\end{example}

\begin{example} {\it A trapezoid, high holonomic rank.}
\rm
The Ore-Sato coefficient $\varphi(s,t)={\rm \Gamma}(s+t){\rm \Gamma}(s)^{k-1}{\rm \Gamma}(-s)^{k}{\rm \Gamma}(-t)$
defines the following hypergeometric system with holonomic rank $k:$
$$
\begin{array}{l}
x \theta_{x}^{k-1} (\theta_x + \theta_y) - (-1)^k \theta_{x}^k, \\
y (\theta_x + \theta_y) + \theta_{y}.
\end{array}
$$
A basis in its solution space is given by $\{ \log^{j}((y+1)/x), \, j=0,\ldots, k-1 \}.$ The generating solution equals $\log^{k-1}((y+1)/x).$ Thus the general solution to this system belongs to~$Cl_1$ by the conservation principle. This example shows that the analytic complexity of solutions to hypergeometric systems with high holonomic rank can still be low.
\label{ex(1,1)(1,0)^(k-1)(-1,0)^k(0,-1)}
\end{example}

\begin{example} {\it A triangle with no symmetries.}
\rm The hypergeometric system
\begin{equation}
\begin{array}{l}
x (\theta_x + \theta_y -4) (\theta_x + 2 \theta_y - 4) -
(2\theta_x + 3 \theta_y - 4) (2\theta_x + 3 \theta_y -
5), \\
y (\theta_x + \theta_y -4) (\theta_x + 2 \theta_y - 4) (\theta_x +
2 \theta_y - 3) \\
\hskip2.3cm - (2\theta_x + 3 \theta_y - 4) (2\theta_x + 3
\theta_y - 5) (2\theta_x + 3 \theta_y - 6)
\end{array}
\label{eq:TriangleWithNoSymmetries}
\end{equation}
is holonomic and its holonomic rank equals~6. The pure basis in its
solution space is given by the Laurent polynomials
$$
x^{-4}y^4,\quad x^{-2}y^3, \quad x^7 y^{-3}, \quad  x^8 y^{-4},
\quad 3 y^2 + 2 x^{-1} y^2,
$$
$$
6 x^2 + 12 x^3 + x^4 + 4 x^5 y^{-2} + 6 x^6 y^{-2} - 12 x^4 y^{-1}
- 4 x^5 y^{-1} - 12 x y - 4 x^2 y.
$$

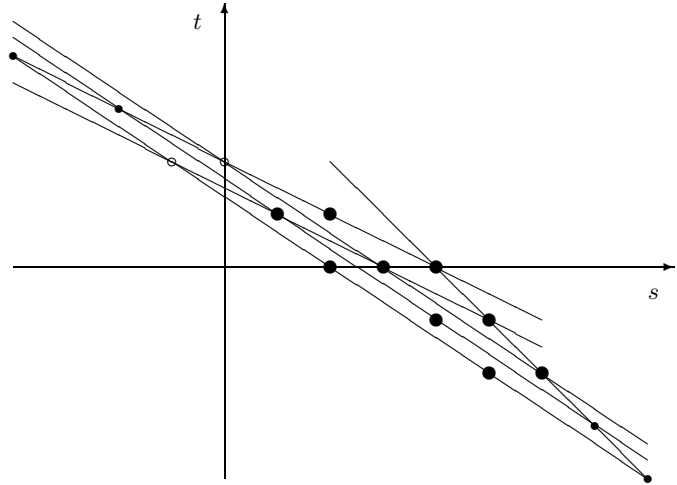
\begin{figure}[htbp]
\centering
\begin{minipage}{7cm}
\begin{picture}(250,180)
  \put(80,0){\vector(0,1){180}}
  \put(0,80){\vector(1,0){250}}
  \put(240,68){\small $s$}
  \put(68,170){\small $t$}
  \put(240,0){\line(-3,2){240}}
  \put(240,7){\line(-3,2){240}}
  \put(240,13){\line(-3,2){240}}
  \put(240,0){\line(-1,1){120}}
  \put(200,60){\line(-2,1){200}}
  \put(200,50){\line(-2,1){200}}
  \put(0,160){\circle*{3}}
  \put(40,140){\circle*{3}}
  \put(220,20){\circle*{3}}
  \put(240,0){\circle*{3}}
  \put(60,120){\circle{3}}
  \put(80,120){\circle{3}}
  \put(100,100){\circle*{5}}
  \put(120,100){\circle*{5}}
  \put(120,80){\circle*{5}}
  \put(140,80){\circle*{5}}
  \put(160,80){\circle*{5}}
  \put(160,60){\circle*{5}}
  \put(180,60){\circle*{5}}
  \put(180,40){\circle*{5}}
  \put(200,40){\circle*{5}}
\end{picture}
\end{minipage}

\caption{The supports of
solutions to the system~(\ref{eq:TriangleWithNoSymmetries})}
\end{figure}

\noindent Here the small filled circles correspond to
monomial solutions, the two empty circles indicate the binomial
solution and the big filled circles correspond to the remaining
polynomial solution. The analytic complexity of the general solution to the system~(\ref{eq:TriangleWithNoSymmetries}) does not exceed~$5.$
\label{ex(1,1)(1,2),(-2,-3)}
\end{example}

\paragraph{Acknowledgements.}

This research was performed in the framework of the state task in the field of scientific activity of the Ministry of Science and Higher Education of the Russian Federation, 
grant no. FSSW-2020-0008.


\begin{thebibliography}{}

\bibitem{Arnold}
Arnold,~V.I.: On the representation of continuous functions of
three variables by superpositions of continuous functions of two
variables. Mat.~Sb.  {\bf 48}(1), 3--74 (1959)

\bibitem{BeloshapkaRJMP2007}
Beloshapka,~V.K.: Analytic complexity of functions of two
variables. Russian J. Math. Phys. {\bf 14}(3), 243--249 (2007)

\bibitem{BeloshapkaRJMP2012}
Beloshapka,~V.K.: Analytical complexity: Development of the topic. 
Russian J. Math. Phys. {\bf 19}(4), 428--439 (2012)


\bibitem{BeloshapkaZametki2019}
Beloshapka,~V.K.: On the complexity of differential algebraic definition for classes of analytic complexity, Math. Notes, {\bf 105} (3), 323--331 (2019) 

\bibitem{DickensteinSadykovAdvances}
Dickenstein,~A., Matusevich,~L.F., Sadykov,~T.M.: Bivariate hypergeometric D-Modules. Advances in Mathematics {\bf 196},
78--123 (2005)

\bibitem{DickensteinSadykovDoklady}
Dickenstein,~A., Sadykov,~T.M.: Algebraicity of~solutions to the
Mellin system and its monodromy. Dokl. Math. {\bf 75}(1),
80--82 (2007)

\bibitem{DickensteinSadykovMatSbornik}
Dickenstein,~A., Sadykov,~T.M.: Bases in the solution space of the Mellin system. Sbornik Mathematics, {\bf 198} (9), 59--80 (2007)

\bibitem{Horn1889}
Horn~J.: \"Uber die Konvergenz der hypergeometrischen Reihen
zweier und dreier Ver\"anderlichen. Math. Ann. {\bf 34}, 544--600 (1889)


\bibitem{KrasikovCASC2019}
Krasikov,~V.A.: Analytic Complexity of Hypergeometric Functions Satisfying Systems with Holonomic Rank Two. Lecture Notes in Computer Science, {\bf 11661}, 330--342 (2019)

%

\bibitem{Sadykov-Tanabe}
Sadykov,~T.M., Tanabe,~S: Maximally reducible monodromy of bivariate hypergeometric systems. Izv.: Math. {\bf 80} (1), 221--262 (2016)

\bibitem{SadykovCASC2018}
Sadykov,~T.M.: Beyond the First Class of Analytic Complexity. Lecture Notes in Computer Science, {\bf 11077}, 335--344 (2018)

\bibitem{SadykovProgramming}
Sadykov,~T.M.: Computational problems of multivariate hypergeometric theory. Programming and Computer Software {\bf 44} (2), 131--137 (2018)

\bibitem{SadykovBullSciMath}
Sadykov,~T.M.: The Hadamard product of hypergeometric series. Bulletin des Sciences Mathematiques {\bf 126} (1), 31 (2002)

\bibitem{SadykovProceedings}
Sadykov,~T.M.: On the analytic complexity of hypergeometric functions. Proceedings of the Steklov Institute of Mathematics {\bf 298} (1), 248--255 (2017)


\bibitem{StepanovaSbMath}
Stepanova,~M.A.: Analytic complexity of differential algebraic functions. Sbornik Mathematics {\bf 210} (12), 1774--1787 (2019) 

\bibitem{StepanovaJSFU}
Stepanova,~M.A.: On analytical complexity of antiderivatives. Journal of Siberian Federal University. Mathematics \& Physics. {\bf 12} (6), 694--698 (2019)

\bibitem{Vitushkin}
Vitushkin,~A.G.: On Hilbert's thirteenth problem and related questions. Russian Math. Surveys, {\bf 59} (1), 11--25 (2004) 

\end{thebibliography}
\end{document}